\documentclass[aps,pre,groupedaddress]{revtex4-2} 
\usepackage{natbib}
\usepackage{amsmath}
\usepackage{epsfig}
\usepackage{color}

\usepackage{graphicx}
\usepackage{amsfonts}
\usepackage{dcolumn}
\usepackage{bm}
\usepackage{color}
\usepackage{hyperref}

\begin{document}

\title{Semi-Markov processes in open quantum systems: Connections and applications in counting statistics}
\author{Fei Liu}
\email[Email address: ]{feiliu@buaa.edu.cn}
\affiliation{School of Physics, Beihang University, Beijing 100191, China}

\date{\today}

\begin{abstract}
Using the age-structure formalism, we definitely establish connections between semi-Markov processes and the dynamics of open quantum systems that satisfy the Markov quantum master equations. A generalized Feynman-Kac formula of the semi-Markov processes is also proposed. In addition to inheriting all statistical properties possessed by the piecewise deterministic processes of wavefunctions, the semi-Markov processes show their unique advantages in quantum counting statistics. Compared with the conventional method of the tilted quantum master equation, they can be applied to more general counting quantities. In particular, the terms involved in the method have precise probability meanings. We use a driven two-level quantum system to exemplify these results.
\end{abstract}
\maketitle

\section{Introduction}
\label{section1}
Stochastic systems, the evolutions of which consist of a mixture of deterministic motion and random jumps, can be modeled as piecewise-deterministic Markov processes (PDPs)~\cite{Davis1984,Davis1993}. The PDPs have a variety of applications in engineering and modeling, e.g., in operations research~\cite{Davis1993} and in modeling biological processes in cells~\cite{Bresslof2014}. In physics, a significant example exists in open quantum systems: it is found that the Markov quantum master equations (MQMEs), which describe the dynamics of the reduced density matrices of the open quantum systems, can be unraveled to the PDPs of the wavefunctions; the individual realizations of these processes are called quantum jump trajectories~\cite{Breuer2002}. From this perspective, a reduced density matrix is equal to a mean of the pure states of the individual quantum systems; the wavefunctions of these systems deterministically evolve in a nonunitary way and are randomly interrupted by collapses. The physical basis behind this approach is quantum measurement theory, in which quantum systems are continuously measured by external detectors~\cite{Mollow1975,Srinivas1981,Mandel1995,Carmichael1993,Plenio1998,Breuer2002,Gardiner2004,
Wiseman2010}. Sophisticated experiments have verified quantum jump trajectories~\cite{Nagourney1986,Bergquist1986,Basche1995,Gleyzes2007,Murch2013,Sun2013,Vool2014,
Campagne-Ibarcq2016,Minev2019,Yan2022}. Unless otherwise stated, the PDPs mentioned in the remainder of this paper always pertain to the wavefunctions of the open quantum systems.

As a well-established notion and useful technique in quantum optics \cite{Molmer93,Scully1997,Gardiner2004}, in the past two decades, PDPs and quantum jump trajectories have also been widely used in the stochastic thermodynamics of open quantum systems~\cite{DeRoeck2006,Esposito2006,Derezinski2008,Crooks2008,Esposito2009,Garrahan2010,Campisi2011,Horowitz2012,Hekking2013,Liu2014a,Liu2016a,Carollo2019,Hasegawa2020,
VanVu2021,Manzano2021,Miller2021,Yada2022,Yan2022}. There are two plausible causes. First, these processes provide a clear picture of measurable trajectories. Hence, extending the classical results based on trajectories~\cite{Seifert2011} to the quantum cases becomes feasible. Second, collapses of the wavefunctions along the quantum jump trajectories enable precise interpretations of energy quanta~\cite{Breuer2003,Horowitz2012,Hekking2013,Liu2018}. This is the key to define thermodynamic quantities and explore thermodynamic laws in the quantum regime.

Among recent applications of the PDPs to the quantum versions ~\cite{Carollo2019,Hasegawa2020,VanVu2021,Liu2021a} of the thermodynamic uncertainty relations~\cite{Barato2015,Gingrich2016,Garrahan2017}, the work of Carollo et al.~\cite{Carollo2019} attracts our attention. They named a certain type of PDPs as quantum rest processes, in which the wavefunctions before and immediately after the collapses are independent, and the collapsed wavefunctions consist of a fixed set. {Carollo et al. argued that these special PDPs are semi-Markov processes (sMPs)}, since the times between collapses are in general nonexponential random variables and are independent of previous history before the last collapses. We note that such PDPs are universal in quantum optics and stochastic thermodynamics, e.g., in the spontaneous fluorescence of two-level atoms~\cite{Mandel1995,Scully1997}. Similar ideas have also existed in the literature for quite a long time~\cite{Zoller1987,Carmichael1989,Mandel1995,Breuer2002}.

Conventional sMPs are mainly concerned with the probabilities of stochastic systems remaining in discrete states and how these quantities evolve with time~\cite{Ross1995,Qian2006,Wang2007}. In contrast, the states or wavefunctions of open quantum systems are constantly changing. To be consistent with quantum dynamics, auxiliary mathematical formalism must be combined into the sMPs. Carollo et al. have solved this problem in special steady states~\cite{Carollo2019}.
In this paper, we attempt to advance this effort and definitely establish connections between sMPs and PDPs in general situations. The other intention of this work is to study the application of sMPs in the counting statistics of open quantum systems~\cite{Srinivas1981,Mandel1995,Levitov1996,Zheng2003,Harbola2006,Esposito2007b,Esposito2009,Bruderer2014}. The latter is a deepening of the former motivation. We will show that the sMPs are not only an alternative mathematic language of the PDPs, but also confer unique advantages in analyzing and computing the counting statistics.

This paper is organized as follows. The first part pertains to the sMPs of classical systems. In Sec.~(\ref{section2}), we briefly review the age structure formalism of the sMPs. Essential notations and formulas are introduced. In Sec.~(\ref{section3}), we propose a generalized Feynman-Kac (FK) formula of the sMPs. Based on the formula, in Sec.~(\ref{section4}), an equation that can calculate the moment generating functions (MGFs) of counting quantities is derived. The second part is fully devoted to the quantum case. In Sec.~(\ref{section5}), after arguing that sMPs exist in open quantum systems, we apply the age structure formalism to reconstruct the dynamics of open quantum systems. In the same section, we prove that the sMPs provide an alternative method to the counting statistics. In Sec.~(\ref{section6}), a driven two-level quantum system exemplifies the previous results. Section~(\ref{section7}) concludes the paper.

\section{Semi-Markov processes}
\label{section2}
We start with the conventional sMPs of the classical systems. The descriptions follow an intuitive age-structure formalism~\cite{Ross1995,Qian2006,Wang2007}. In Sec.~\ref{section5}, we will show that this theory is able to reconstruct the dynamics of open quantum systems. Let $p_{\alpha|\beta}(\tau)$ be the waiting time density of a sMP, i.e., the probability density of jumping out of state $\alpha$ to $\beta$ at age $\tau$ since the system arrival to state $\alpha$. The states of the classical system are thought to be discrete and finite. The survival distribution function $S_\alpha(\tau)$ is the probability of the system remaining in state $\alpha$ without jumps until age $\tau$. These probabilities are connected by
\begin{eqnarray}
\label{survivalprob1}
S_\alpha(\tau)=\int_\tau^\infty \left[\sum_{\beta\ne\alpha}p_{\alpha|\beta}(\tau')\right]d\tau'.
\end{eqnarray}
It is very useful to introduce the hazard function $k_{\alpha\beta}(\tau)$, which satisfies
\begin{eqnarray}
\label{differentialsurvivalprob}
dS_\alpha(\tau)=-S_\alpha(\tau)\left[\sum_{\beta\neq\alpha} k_{\alpha|\beta}(\tau)\right]d\tau\equiv-S_\alpha(\tau)\Gamma_\alpha(\tau)d\tau.
\end{eqnarray}
This equation indicates the probability mean of these functions: they include the conditional probability density of jumping out of state $\alpha$ to $\beta$ at age $\tau$, while $\Gamma_\alpha(\tau)$ is the total conditional probability density of jumping out of state $\alpha$. Comparing Eq.~(\ref{survivalprob1}) with (\ref{differentialsurvivalprob}), we see that the waiting time density and survival distribution can be rewritten by the hazard functions as
\begin{eqnarray}
p_{\alpha|\beta}(\tau)=S_{\alpha}(\tau)k_{\alpha|\beta}(\tau)
\end{eqnarray}
and
\begin{eqnarray}
\label{survivalprob2}
S_\alpha(\tau)=e^{-\int_0^\tau \Gamma_{\alpha}(\tau')d\tau'},
\end{eqnarray}
respectively.

Let $p_\alpha(t,\tau)$ be the probability density of the system in state $\alpha$ at time $t$ with age $\tau$. The evolution equation of the density is~\cite{Qian2006,Wang2007,Ross1995}
\begin{eqnarray}
\label{evoeqdoubletimeprob}
\partial_t p_\alpha(t,\tau)+\partial_\tau p_\alpha(t,\tau)=-\Gamma_\alpha(\tau) p_\alpha(t,\tau).
\end{eqnarray}
Eq.~(\ref{evoeqdoubletimeprob}) is obtained by expanding $p_\alpha(t+h,\tau+h)=[1-\Gamma_\alpha(\tau)h]p_\alpha(t,\tau)$ until the first order of the small time interval $h$. Note that at age zero,
\begin{eqnarray}
\label{doubletimeprobage0}
p_\alpha(t,0)=\sum_{\beta\ne\alpha}\int_0^t p_\beta(t,\tau)k_{\beta|\alpha}(\tau)d\tau+\delta_{\alpha\gamma}\delta(t).
\end{eqnarray}
Here, for simplicity, we have stipulated that at time $t=0$, the system always departs from state $\gamma$ with age zero. With Eqs.~(\ref{evoeqdoubletimeprob}) and~(\ref{doubletimeprobage0}), the probability density of the system in state $\alpha$ at time $t$, is
\begin{eqnarray}
\label{probatdiscretestate}
p_\alpha(t)=\int_0^t p_\alpha(t,\tau)d\tau,
\end{eqnarray}
which satisfies the generalized master equation (GME)
\begin{eqnarray}
\label{GME}
\frac{d}{dt}p_\alpha(t)=\sum_{\beta\ne\alpha} \left[ p_\beta(t)*K_{\beta|\alpha}(t)- p_\alpha(t)*K_{\alpha|\beta}(t) \right],
\end{eqnarray}
and the initial density $p_\alpha(0)$ equals $\delta_{\alpha\gamma}$. In Eq.~(\ref{GME}), the asterisks represent convolutions, and the memory kernel $K_{\alpha|\beta}(t)$ is an inverse Laplace transform of
\begin{eqnarray}
\hat{K}_{\alpha|\beta}(v)=\frac{\hat{p}_{\alpha|\beta}(v)}{\hat{S}_\alpha(v)}.
\end{eqnarray}
Unless otherwise stated, the marks ( $\hat{ }$ ) placed over symbols denote Laplace transforms, e.g.,
\begin{eqnarray}
\hat{p}_{\alpha|\beta}(v)\equiv {\cal L}[p_{\alpha|\beta}(\tau)]
=\int_0^\infty e^{-vs}p_{\alpha|\beta}(s)ds.
\end{eqnarray}
Here, we do not explain the details of the GME~\cite{Qian2006,Wang2007}.

\section{Generalized Feynman-Kac formula of semi-Markov process}
\label{section3}
Individual realizations of the sMPs are named trajectories. Using the waiting time densities and survival distributions, we write the probability density of observing a trajectory $X$ as
\begin{eqnarray}
\label{probdistributiontraj}
{\cal P}[X]
=p_{\alpha_0|\alpha_1}(\tau_1)p_{\alpha_1|\alpha_2}(\tau_2)\cdots p_{\alpha_{N-1}|\alpha_N}(\tau_N)S_{\alpha_N}(\tau_{N+1}).
\end{eqnarray}
In the trajectory, we have assumed that there are a total of $N$ jumps of the state of the system occurring at times $t_i$. The age $\tau_i$ is equal to $t_i-t_{i-1}$, and $\tau_{N+1}=t-t_N$. The duration of the process is set to $t$. In addition, the states before and immediately after the $i$-th jump are $\alpha_{i-1}$ and $\alpha_i$, respectively, $i=0,1,\cdots, N$.

Consider a random functional of the trajectory $X$:
\begin{eqnarray}
\label{generalrandomfunctional}
A[X]=\int_0^{\tau_1} V_{\alpha_0}(\tau')d\tau' +\cdots +\int_0^{\tau_N} V_{\alpha_{N-1}}(\tau') d\tau' +\int_0^{\tau_{N+1}} V_{\alpha_N}(\tau') d\tau',
\end{eqnarray}
where $V_\alpha(\tau)$ is an arbitrary function of state $\alpha$ and age $\tau$. These functions are thought to be continuous with respect to the age variable. We are interested in the probability density $p_A(u)$ of the random variable~(\ref{generalrandomfunctional}). A conventional routine is to compute the MGF and to then conduct an inverse Laplace transform. The former is
\begin{eqnarray}
\label{momentgeneratingfungeneralFK}
\Phi(\eta,t)=\int e^{-\eta u}p_A(u)du=\left\langle e^{-\eta A[X]}\right\rangle,
\end{eqnarray}
where the angular brackets denote an average over all possible trajectories of the sMP.

At first sight, the MGF~(\ref{momentgeneratingfungeneralFK}) seems useless due to the unknown $p_A(u)$. Nevertheless, we follow Kac's idea~\cite{Kac1949} to prove that this quantity can be solved by a differential equation such as the GME~(\ref{GME}). To this end, let $p_\alpha(u,t,\tau)$ be the joint probability density of the system in state $\alpha$ with age $\tau$, where the random variable $A$ equals $u$ at time $t$ simultaneously. The evolution equation of this density is derived by carrying out a similar argument as in Eqs.~(\ref{evoeqdoubletimeprob}) and~(\ref{doubletimeprobage0}):
\begin{eqnarray}
\label{evoeqdoubletimeprobfunctionalFK}
&&\partial_t p_\alpha(u,t,\tau)+\partial_\tau p_\alpha(u, t,\tau)=-\Gamma_\alpha(\tau) p_\alpha(u,t,\tau)- V_\alpha(\tau)\partial_u p_\alpha(u,t,\tau),
\\
\label{initialconditionFK}
&&p_\alpha(u,t,0)=\sum_{\beta\ne\alpha}\int_0^t p_\beta(u,t,\tau)k_{\beta|\alpha}(\tau)d\tau +\delta_{\alpha\gamma}\delta(t)\delta(u).
\end{eqnarray}
With the joint probability density, Eq.~(\ref{momentgeneratingfungeneralFK}) is rewritten as
\begin{eqnarray}
\label{q(alpha,eta,t,tau)function}
\Phi(\eta,t)&=&\int_0^t\int e^{-\eta u}\sum_\alpha p_\alpha(u,t,\tau)du d\tau \nonumber\\
&\equiv &\int_0^t \sum_\alpha q_\alpha(\eta,t,\tau)d\tau\equiv \sum_\alpha P_\alpha(\eta,t).
\end{eqnarray}
In the second and third equations, $q_\alpha(\eta,t,\tau)$ and $P_\alpha(\eta,t)$ are defined. Eqs.~(\ref{evoeqdoubletimeprobfunctionalFK}) and~(\ref{initialconditionFK}) lead to two equations:
\begin{eqnarray}
\label{evoeqdoubletimeprobfunctionalFK2}
&&\partial_t q_\alpha(\eta,t,\tau)+\partial_\tau q_\alpha(\eta, t,\tau)=-\Gamma_\alpha(\tau) q_\alpha(\eta,t,\tau)- \eta V_\alpha(\tau)  q_\alpha(\eta,t,\tau),\\
\label{initialconditionFK2}
&&q_\alpha(\eta,t,0)=\sum_{\beta\ne\alpha}\int_0^t q_\beta(\eta,t,\tau)k_{\beta|\alpha}(\tau)d\tau +\delta_{\alpha\gamma}\delta(t).
\end{eqnarray}
For Eq.~(\ref{evoeqdoubletimeprobfunctionalFK2}), there is a formally exact solution:
\begin{eqnarray}
\label{qsolution}
q_\alpha(\eta,t,\tau)=S_\alpha(\tau)e^{-\eta\int_0^\tau V_\alpha(\tau')d\tau'}q_\alpha(\eta,t-\tau,0).
\end{eqnarray}
Integrating it over time $\tau$ and substituting the result into Eq.~(\ref{initialconditionFK2}), we have
\begin{eqnarray}
\label{Pfunc}
&&P_\alpha(\eta,t)=\overline{S}_\alpha(\eta,t)*q_\alpha(\eta,t,0),\\
\label{initialconditionFK3}
&&q_\alpha(\eta,t,0)=\sum_{\beta\ne\alpha}\overline{p}_{\beta|\alpha}(\eta,t)*q_\beta(\eta,t,0) +\delta_{\alpha\gamma}\delta(t),
\end{eqnarray}
where
\begin{eqnarray}
\label{survivalprobrevisedFK}
\overline{S}_\alpha(\eta,\tau)&=&S_\alpha(\tau)e^{-\eta\int_0^\tau V_\alpha(\tau')d\tau'},\\
\label{sMPkernelrevisedFK}
\overline{p}_{\alpha|\beta}(\eta,\tau)
&=&p_{\alpha|\beta}(\tau)e^{-\eta\int_0^\tau V_\alpha(\tau')d\tau'}
=\overline{S}_\alpha(\eta,\tau)k_{\alpha|\beta}(\tau).
\end{eqnarray}
The final step is to apply the Laplace transform of time $t$ in Eqs.~(\ref{Pfunc}) and (\ref{initialconditionFK3}) and eliminate $\hat q_\alpha(\eta,v)\equiv{\cal L}[ q_\alpha(\eta,t,0)]$. We have
\begin{eqnarray}
\label{FKsMPs}
v \hat P_\alpha(\eta,v)-\delta_{\alpha\gamma}&=&\sum_{\beta\neq\alpha} \left[\hat P_\beta(\eta,v)\frac{\hat{ \overline{p}}_{\beta|\alpha}(\eta,v)}{\hat{\overline{S}}_\beta(\eta,v)}-\hat P_\alpha(\eta,v)\frac{\hat{\overline{p}}_{\alpha|\beta}(\eta,v)}{\hat{\overline{S}}_\alpha(\eta,v)}\right]
-\frac{\eta}{2\pi{\rm i}}\frac{\hat V_\alpha(v)* \hat{\overline{S}}_\alpha(\eta,t) }{\hat{\overline{S}}_\alpha(\eta,v)} \hat P_\alpha(\eta,v).
\end{eqnarray}
Hence, if these algebraic equations are solved, we will obtain the MGF $\Phi(\eta,t)$ by taking an inverse Laplace transform of $\hat \Phi(\eta,v)=\sum_
\alpha \hat P_\alpha(\eta,v)$.

We name Eq.~(\ref{FKsMPs}) the generalized FK formula of the sMPs. The cause is as follows. If $V_\alpha(\tau)$ is independent of age $\tau$, and the memory kernels are proportional to the Dirac functions, i.e., $K_{\alpha|\beta}(t)=2k_{\alpha|\beta}\delta(t)$,
the inverse Laplace transform of the generalized FK formula is
\begin{eqnarray}
\label{standardFK}
\frac{d}{dt}P_\alpha(\eta,t)&=&\sum_{\beta\ne\alpha}\left[ P_\beta(\eta,t)k_{\beta|\alpha} - P_\alpha(\eta,t) k_{\alpha|\beta}\right]-\eta V_\alpha  P_\alpha(\eta,t).
\end{eqnarray}
This is nothing but the canonical FK formula of Markov jump processes~\cite{Imparato2007,Lebowitz1999}. It is worth pointing out that Eqs.~(\ref{evoeqdoubletimeprobfunctionalFK2})-(\ref{FKsMPs}) also account for the derivation of the GME: we set the parameter $\eta$ to zero; then, $\overline{S}_\alpha(\eta,\tau)\rightarrow S_\alpha(\tau)$ and $\overline{p}_{\alpha|\beta}(\eta,\tau)\rightarrow p_{\alpha|\beta}(\tau)$, and the inverse Laplace transform of Eq.~(\ref{FKsMPs}) leads to Eq.~(\ref{GME}).


\section{Counting statistics of semi-Markov processes}
\label{section4}
It is of interest to study the counting statistics of random quantities such as
\begin{eqnarray}
\label{countingnumber}
Q[X]=\sum_{i=1}^N \omega_{\alpha_{i-1}\alpha_i},
\end{eqnarray}
$\omega_{\alpha_{i-1}\alpha_i}$ denotes an arbitrary weight specified by the states immediately before and immediately after the $i$-th jump. The simplest case is that all the weights are equal to $1$. Then, $Q[X]$ is equal to the total number of jumps along the trajectory $X$. Earlier work has studied the counting statistics of sMPs, e.g., the fluctuation theorems~\cite{Andrieux2008,Esposito2007}. In particular, an equation analogous to Eq.~(\ref{GME}) was obtained to calculate the MGF of the random variable~(\ref{countingnumber})~\cite{Cavallaro2016,Esposito2007}. Here, we show that the previous result is a special case of the generalized FK formula~(\ref{FKsMPs}). We must emphasize that our goal is not only to propose an alternative way of deriving the same equation; what truly matters is the age-structure formalisms behind Eqs.~(\ref{evoeqdoubletimeprobfunctionalFK}) and~(\ref{initialconditionFK}).

Let the MGF of the random variable Eq.~(\ref{countingnumber}) be
\begin{eqnarray}
\label{momentgeneratingfuncofcounting}
M(\lambda,t)
=\left\langle e^{-\lambda Q[X]}\right\rangle.
\end{eqnarray}
Eqs.~(\ref{countingnumber}) and~(\ref{generalrandomfunctional}) appear different. To this end, we write Eq.~(\ref{momentgeneratingfuncofcounting}) as an explicit expression by substituting Eqs.~(\ref{probdistributiontraj}) and (\ref{countingnumber}):
\begin{eqnarray}
\label{momentgenertingfuncofcounting2}
M(\lambda,t)&=&\sum_X k_{\alpha_0|\alpha_1}(\tau_1)e^{-\lambda\omega_{\alpha_0|\alpha_1}}S_{\alpha_0}(\tau_1)\cdots k_{\alpha_{N-1}|\alpha_N}(\tau_N)e^{-\lambda\omega_{\alpha_{N-1}|\alpha_N}}S_{\alpha_{N-1}}(\tau_N)S_{\alpha_N}(t-t_N)
\nonumber\\
&=&\sum_X k'_{\alpha_0|\alpha_1}(\tau_1)S'_{\alpha_0}(\tau_1)\cdots k_{\alpha_{N-1}|\alpha_N}(\tau_N)S'_{\alpha_{N-1}}(\tau_N)S'_{\alpha_N}(t-t_N)e^{\sum_{i=1}^{N+1}\int_0^{\tau_i} [ \Gamma_{i-1}(\tau')-\Gamma'_{i-1}(\tau')]d\tau'}\nonumber \\
&\equiv& \left\langle e^{\sum_{i=1}^{N+1}  \int_0^{\tau_i}V_{\alpha_{i-1}}(\tau') ]d\tau'}\right\rangle'.
\end{eqnarray}
In the third equation, we use the mark ($'$) to denote another sMP with modified hazard functions
\begin{eqnarray}
\label{newhazardfuncs}
k'_{\alpha|\beta}(\tau)=k_{\alpha|\beta}(\tau)e^{-\lambda \omega_{\alpha \beta}}.
\end{eqnarray}
Accordingly, the survival distribution $S'_\alpha(\tau)$ is similar to Eq.~(\ref{survivalprob2}), except that $\Gamma(\tau)$ therein is replaced by $\Gamma'_\alpha(\tau)=\sum_{\beta\neq\alpha } k'_{\alpha|\beta}(\tau)$. Obviously,
\begin{eqnarray}
\label{countingfunctional}
V_\alpha(\tau)=\Gamma'_\alpha(\tau)-\Gamma_\alpha(\tau).
\end{eqnarray}

When we compare Eqs.~(\ref{momentgenertingfuncofcounting2}) with~(\ref{momentgeneratingfungeneralFK}), we see that the generalized FK formula is available at this point. The former is equal to the latter, with $\eta=-1$ and $V_\alpha(\tau)$ defined in Eq.~(\ref{countingfunctional}). The reader is reminded that now the hazard functions of the new sMP are Eq.~(\ref{newhazardfuncs}). In this situation, Eqs.~(\ref{survivalprobrevisedFK}) and~(\ref{sMPkernelrevisedFK}) are simply
\begin{eqnarray}
\overline{S}_\alpha(\eta=-1,\tau)&=&S_\alpha(\tau),\\
\overline{p}_{\alpha|\beta}(\eta=-1,\tau)&=&p_{\alpha|\beta}(\tau)e^{-\lambda\omega_{\alpha\beta} }.
\end{eqnarray}
Substituting all the results into Eq.~(\ref{FKsMPs}), we immediately have
\begin{eqnarray}
\label{MEofcountingstatistics}
v \hat P_{\alpha}(v)-\delta_{\alpha\gamma}&=&\sum_{\beta\neq\alpha} \left[\hat P_{\beta}(v)\frac{\hat{p}_{\beta|\alpha}(v)}{\hat{S}_\beta(v)}e^{-\lambda\omega_{\beta\alpha}}-\hat P_{\alpha}(v)\frac{\hat{p}_{\alpha|\beta}(v)}{\hat{S}_\alpha(v)}\right].
\end{eqnarray}
Note that the parameter $\eta$ is abandoned because it is equal to $-1$. After solving the algebraic equations, we can calculate Eq.~(\ref{momentgenertingfuncofcounting2}) by taking an inverse Laplace transform of $\hat M(\lambda,\eta)=\sum_{\alpha} \hat P_{\alpha}(v)$. Before closing the discussions about the sMPs of the classical systems, let us mention that if taking inverse Laplace transforms on both sides of Eq.~(\ref{MEofcountingstatistics}), an equation analogous to the GME will be obtained; the only difference is that in the former, there is an additional term $\exp(-\lambda\omega_{\alpha\beta})$ in front of the first term on the right-hand side of Eq.~(\ref{GME}).

\section{Semi-Markov processes in open quantum systems}
\label{section5}
Before we expound the sMPs in the open quantum systems, we first sketch the MQME and its unraveling of the PDPs~\cite{Breuer2002}. Let $\rho(t)$ be the reduced density matrix of an open quantum system. Under appropriate assumptions and conditions, the dynamics of the system is described by MQME~\cite{Davies1974,Lindblad1976,Gorini1976}:
\begin{eqnarray}
\label{QME}
\partial_t \rho(t)=-{\rm i}[H,\rho(t)]+\sum_{\alpha=1}^M r_\alpha\left( A_\alpha\rho(t)A^\dag_\alpha -\frac{1}{2}\left\{A^\dag_\alpha A_\alpha,\rho(t)\right\}\right),
\end{eqnarray}
where the Planck constant $\hbar$ is set to 1, $H$ denotes the Hamiltonian of the quantum system, $A_\alpha$ is the Lindblad operator, and the nonnegative $r_\alpha$, $\alpha=1,\cdots,M$, represent the correlation functions of the environment surrounding the system. Eq.~(\ref{QME}) can be unraveled into the PDP~\cite{Breuer2002}, and the individual realizations of the process are the quantum jump trajectories~\cite{Breuer2002,Gardiner2004,Wiseman2010,Plenio1998,Carmichael1993}. These trajectories, which pertain to the evolutions of the wavefunctions of the single quantum systems, are composed of deterministic pieces and random collapses of the wavefunctions. The former are the solutions of the nonlinear Schr$\ddot{o}$dinger equation,
\begin{eqnarray}
\label{nonlinearSchrodingerequation}
\frac{d}{d\tau}\psi(\tau)  &=&-{\rm i}{ G}[\psi(\tau)],
\end{eqnarray}
where $\tau$ is set to zero immediately after the last collapse, and the operator $G$ is
\begin{eqnarray}
\label{Goperator}
{ G}[\psi]=\left(\hat H  + \frac{{\rm i}}{2}\sum_{\alpha=1}^M r_\alpha \parallel A_\alpha\psi\parallel ^2\right ) \psi,
\end{eqnarray}
and the non-Hermitian Hamiltonian $\hat H$ is equal to $H-({\rm i}/{2})\sum_{\alpha=1}^M r_\alpha A_\alpha^\dag A_\alpha $. The latter collapses are
\begin{eqnarray}
\label{targetwavefunction}
\psi(\tau)  \rightarrow \phi_\alpha=\frac{A_\alpha \psi(\tau) }{\parallel A_\alpha \psi(\tau) \parallel},
\end{eqnarray}
and the rates of the collapses are
\begin{eqnarray}
\label{agerates}
w[\psi(\tau)|\phi_\alpha]=r_\alpha \parallel A_\alpha|\psi(\tau)\rangle \parallel ^2,
\end{eqnarray}
$\alpha=1,\cdots,M$. Quantum jump trajectories are present if the quantum systems are continuously measured or monitored~\cite{Wiseman2010,Breuer2002}.

\begin{figure}
\includegraphics[width=1.\columnwidth]{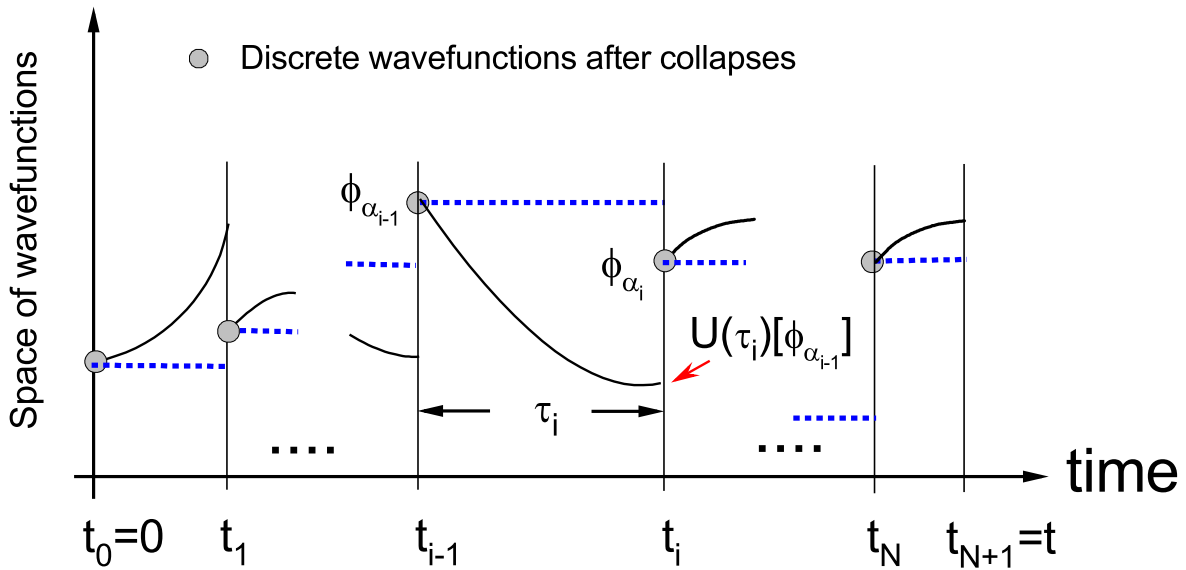}
\caption{Schematic diagrams of a quantum jump trajectory of the PDP and a trajectory of the sMP in an open quantum system. The curves represent the former. If we are only concerned about the wavefunctions after collapses, $\phi_{\alpha_i}$, which are represented by the gray dots, we imagine them as trajectories of the sMP embedded in the PDP. As a comparison, we also plot a trajectory of a conventional sMP; see the dotted horizontal lines. Before any jumps occur, the system will remain in the state at the time of departure. }
\label{fig1}
\end{figure}

When we compare the PDP with the sMP in Sec.~\ref{section2}, we observe that there is a sMP embedded in the former only if the instantaneous collapses of the wavefunctions are concerned; time $\tau$ in Eq.~(\ref{nonlinearSchrodingerequation}) plays the role of age in the sMPs. This observation becomes more obvious if the collapsed wavefunctions $\phi_\alpha$ are age-independent and come from a fixed set with finite elements. Carollo et al.~\cite{Carollo2019} named these special PDPs stochastic reset processes. These processes cannot cover all cases, but they are adequate in the most physically interesting situations~\cite{Breuer2002}. Hence, we still refer to them as the PDPs for simplicity. Fig.~(\ref{fig1}) presents schematic diagrams of a quantum jump trajectory and a trajectory of the embedded sMP in an open quantum system.

Now, let us collect the key quantities of the sMPs of the open quantum systems. The first is the hazard function,
\begin{eqnarray}
\label{hazardfuncembeddedsMP}
k_{\alpha|\beta}(\tau)=r_\beta\parallel A_\beta U(\tau)[\phi_\alpha] \parallel^2,
\end{eqnarray}
where $\alpha,\beta=1,\cdots,M$, and $U(t)$ is the nonlinear time-evolution operator of Eq.~(\ref{nonlinearSchrodingerequation}). The cause is simple: the initial condition of Eq.~(\ref{nonlinearSchrodingerequation}) is also the collapsed wavefunction, which we consider as one of the wavefunctions of the set. In addition, the survival distribution and waiting time density are
\begin{eqnarray}
\label{nonlineartimeevolutionoperator}
S_\alpha(\tau)=\parallel e^{-{\rm i}\tau \hat H}\phi_\alpha\parallel^2,
\end{eqnarray}
and
\begin{eqnarray}
\label{waitingtimedensityquantum}
p_{\alpha|\beta}(\tau)=r_\beta\parallel A_\beta e^{-{\rm i}\tau \hat H}\phi_\alpha \parallel^2,
\end{eqnarray}
respectively. These three formulas are known in the Monte Carlo simulations of the quantum jump trajectories~\cite{Breuer2002}. Recall that the indices $\alpha$ and $\beta$ in Eqs.~(\ref{hazardfuncembeddedsMP}) and~(\ref{waitingtimedensityquantum}) may be the same. This point is significant in contrast to the conventional sMPs in the classical regime.

\subsection{Reconstruction of the Markov quantum master equation}
As we mentioned in Sec.~(\ref{section1}), the sMPs alone cannot reconstruct the quantum dynamics of the open quantum systems; an auxiliary mathematical structure is needed. To this end, we propose a relation:
\begin{eqnarray}
\label{connectionofprobsMP2PDP}
P[\psi,t]=\sum_{\alpha=1}^M\int_0^t p_\alpha(t,\tau)\delta\left[\psi-U(\tau)[\phi_\alpha] \right]d\tau,
\end{eqnarray}
where $P[\psi,t]$ is the probability distribution functional of the random wavefunction $\psi$ at time $t$, and $\delta[$ $]$ is the Dirac functional. The probability mean of Eq.~(\ref{connectionofprobsMP2PDP}) is intuitive. Taking the time partial derivative of $P[\psi,t]$ and substituting Eq.~(\ref{evoeqdoubletimeprob}), we have
\begin{eqnarray}
\label{connectionofprobsMP2PDPdetail}
\partial_t P[\psi,t]&=&\sum_{\alpha=1}^M p_\alpha(t,0)\delta\left[\psi-\phi_\alpha \right]+\sum_{\alpha=1}^M\int_0^t  p_\alpha(t,\tau)\partial_\tau\delta\left[\psi-U(\tau)[\phi_\alpha] \right]d\tau\nonumber\\
&&-\sum_{\alpha=1}^M\int_0^t\Gamma_\alpha(\tau) p_\alpha(t,\tau)  \delta\left[\psi-U(\tau)[\phi_\alpha] \right]d\tau.
\end{eqnarray}
Here, integration of parts has been employed. The next step is to substitute Eq.~(\ref{doubletimeprobage0}) and to take the time derivative of the Dirac functional. A calculation shows that
\begin{eqnarray}
\label{quantumLiouvillemasterequation}
\partial_t P[\psi,t]&=&i\int dz\left ( \frac{\delta}{\delta \psi(z)}G[\psi](z)-\frac{\delta}{\delta \psi^*(z)}G[\psi]^*(z) \right ) P[\psi,t]\nonumber +\nonumber \\
&&\int D\phi D\phi^*\left ( P[\phi,t]W[\phi|\psi] -P[\psi,t]W[\psi|\phi] \right ),
\end{eqnarray}
where $\delta/\delta \psi(z)$ and $\delta/\delta^* \psi(z)$ are functional derivatives, $z$ denotes the positional coordinate,
\begin{eqnarray}
\label{totalrate1}
W[\phi|\psi]=\sum_{\alpha=1}^M w[\phi|\phi_\alpha] \delta\left[\phi_\alpha  -\psi\right],
\end{eqnarray}
and the rate $w[\psi|\phi_\alpha]$ is given in Eq.~(\ref{agerates}). Because the calculation is simple, we do not show it here. Eq.~(\ref{quantumLiouvillemasterequation}) is the Liouville-master equation of the PDPs in the Hilbert space~\cite{Breuer2002}. In principle, this equation completely describes the quantum dynamics of open quantum systems. On the other hand, in practice, the MQME~(\ref{QME}) will be more familiar and useful. These equations can be connected by the following equation~\cite{Breuer1995a,Breuer1995,Breuer1997,Breuer1997a,Breuer2002}:
\begin{eqnarray}
\label{densitymatrixzz}
\rho(t)&=&\int D\psi D\psi^* P[\psi,t] |\psi\rangle\langle \psi|,
\end{eqnarray}
where $D\psi D\psi^*$ represents the Hilbert space volume element. We do not explain the details of this connection. However, we find that a combination of Eqs.~(\ref{connectionofprobsMP2PDP}) and~(\ref{densitymatrixzz}) suggests an alternative form of the reduced density matrix:
\begin{eqnarray}
\label{densitymatrixzz2}
\rho(t)&=&\sum_{\alpha=1}^M\int_0^t p_\alpha(t,\tau)  |U(\tau)[\phi_\alpha]\rangle\langle U(\tau)[\phi_\alpha]|. 
\end{eqnarray}
Taking its time derivative and using Eq.~(\ref{evoeqdoubletimeprob}) and~(\ref{doubletimeprobage0}), we can derive MQME~(\ref{quantumLiouvillemasterequation}) in a direct and efficient way. The details are presented in Appendix~\ref{AppendixA}.

Eq.~(\ref{densitymatrixzz2}) implies an intriguing consequence. Assuming that Eq.~(\ref{evoeqdoubletimeprob}) has a stationary solution, that is, when the duration $t$ is long, the probability density~\cite{Wang2007,Ross1995}
\begin{eqnarray}
\label{stationaryagesolution}
p_\alpha(t=\infty,\tau)=c_\alpha S_\alpha(\tau),
\end{eqnarray}
where the coefficient
\begin{eqnarray}
c_\alpha=\frac{\pi_\alpha}{\sum_{\beta=1}^M \pi_\beta\tau_\beta},
\end{eqnarray}
time $\tau_\beta$ is the average age of the system starting from the wavefunction $\phi_\alpha$, i.e.,
\begin{eqnarray}
\tau_\beta=\int_0^\infty {\rm d}\tau S_\beta(\tau),
\end{eqnarray}
and
\begin{eqnarray}
\label{eigenvector}
\pi_\alpha&=&\sum_{\beta=1}^M \pi_\beta \left[ \int_0^\infty {\rm d}\tau p_{\beta|\alpha}(\tau)\right] \equiv  \sum_{\beta=1}^M \pi_\beta P_{\beta|\alpha }.
\end{eqnarray}
Eq.~(\ref{eigenvector}) indicates that $\pi_\alpha$ is the stationary distribution of a Markov chain with transition probability $P_{\alpha|\beta}$, $\alpha,\beta=1,\cdots,M$. It is not difficult to deduce that $c_\alpha$ is the stationary rate of the quantum system collapsed to or departing from the wavefunction $\phi_\alpha$. Substituting Eq.~(\ref{stationaryagesolution}) into~(\ref{densitymatrixzz2}), we have the steady-state solution of MQME~\cite{Liu2021a}:
\begin{eqnarray}
\label{steadystaterhoprobform}
\rho(t=\infty)=\sum_{\alpha=1}^M c_\alpha \int_0^\infty  e^{-i\tau\hat H}|\phi_\alpha \rangle\langle \phi_\alpha |e^{i\tau\hat H^\dag}d\tau.
\end{eqnarray}
Eq.~(\ref{steadystaterhoprobform}) clearly indicates that the reduced density matrix is an incoherent superposition of various pure states.

\subsection{Reconstruction of the tilted quantum master equation}
The counting statistics of the collapses of the wavefunctions serve important roles in quantum optics and quantum stochastic thermodynamics~\cite{Mandel1995,Esposito2009}. A foundational example of the former is that a photon counter continuously detects the fluorescence photons emitted by a two-level atom~\cite{Srinivas1981,Mollow1975,Carmichael1989,Wiseman1993,Mandel1995}. In the latter, the random collapses along the quantum jump trajectories indicate that discrete amounts of energy quanta are released to or absorbed from the environment~\cite{Breuer2003,Derezinski2008,Crooks2008, Campisi2011,Horowitz2012,Hekking2013,Liu2016a,Carollo2019,Hasegawa2020, Manzano2021,Miller2021,Yada2022,Yan2022}. Although these quantities still follow the general Eq.~(\ref{countingnumber}), they usually have a more unique form in the quantum regime:
\begin{eqnarray}
\label{quantumcountingnumber}
Q[X]&=&\sum_{i=1}^N \omega_{\alpha_i}.
\end{eqnarray}
In other words, the arbitrary weight $\omega_{\alpha_i}$ is solely determined by the collapsed wavefunction $\phi_{\alpha_i}$ instead of both $\phi_{\alpha_{i-1}}$ and $\phi_{\alpha_i}$. Of course, this form includes the case of constant weights. Undoubtedly, the counting statistics of the sMPs in Sec.~\ref{section4} also holds in the quantum case. On the other hand, a very influential method of studying counting statistics is the tilted quantum master equation (TQME)~\cite{Mollow1975,Zoller1987,Carmichael1989,Levitov1996,Zheng2003,Esposito2006,Esposito2009,
Garrahan2010,Bruderer2014,Liu2016a}~\footnote{Because the method was developed by different authors in different situations, various names were used, e.g., a hierarchy of quantum master equation~\cite{Carmichael1989,Zheng2003}, the biased quantum master equation~\cite{Bruderer2014}, and the generalized quantum master equation~\cite{Esposito2009}, etc.}. We naturally conclude that these two methods are equivalent for the {\it quantum} counting statistics.

Here, we make use of the age structure formalism to prove this expected equivalence. Let the MGF of Eq.~(\ref{quantumcountingnumber}) be $M(\lambda,t)$. In the current situation, Eq.~(\ref{momentgeneratingfuncofcounting}) and~(\ref{MEofcountingstatistics}) are valid, and $\omega_{\beta\alpha}$ in the latter is replaced by $\omega_\alpha$. Accordingly, Eqs.~(\ref{evoeqdoubletimeprobfunctionalFK2}) and~(\ref{initialconditionFK2}) are simplified to
\begin{eqnarray}
\label{evoeqdoubletimeprobfunctionalFK4}
&&\partial_t q_\alpha(t,\tau)+\partial_\tau q_\alpha(t,\tau)=-\Gamma_\alpha(\tau) q_\alpha(t,\tau),\\
\label{initialconditionFK4}
&&q_\alpha(t,0)=\sum_{\beta}\int_0^t q_\beta(t,\tau)k'_{\beta|\alpha}(\tau)d\tau +\delta_{\alpha\gamma}\delta(t),
\end{eqnarray}
and $k'_{\alpha|\beta}=k_{\alpha|\beta}\exp(-\lambda\omega_\beta)$. We have abandoned the parameter $\eta=-1$. Inspired by Eq.~(\ref{connectionofprobsMP2PDP}), we propose a functional
\begin{eqnarray}
\label{connectionofcountingsMP2PDP}
\widetilde{P}[\psi,t]=\sum_{\alpha=1}^M\int_0^t q_\alpha(t,\tau)\delta\left[\psi-U(\tau)[\phi_\alpha]\right]d\tau.
\end{eqnarray}
Taking the time partial derivative of $P[\psi,t]$, substituting Eq.~(\ref{evoeqdoubletimeprobfunctionalFK4}) and~(\ref{initialconditionFK4}), and carrying out similar calculations as in Eq.~(\ref{connectionofprobsMP2PDPdetail}), we have
\begin{eqnarray}
\label{quantumtiltinggeneratorlevel1}
\partial_t \widetilde P [\psi,t]&=&\text{i}\int dx\left ( \frac{\delta}{\delta \psi(x)}G[\psi](x)-\frac{\delta}{\delta \psi^*(x)}G[\psi]^*(x) \right )\widetilde  P [\psi,t] \nonumber \nonumber \\
&&+\int D\phi D\phi^* \left ( \widetilde P [\phi,t]W'[\phi|\psi]-\widetilde P  [\psi,t]W[\psi|\phi] \right ),
\end{eqnarray}
where
\begin{eqnarray}
\label{rateauxiliarysystem}
W'[\phi|\psi]=\sum_{\alpha=1}^M \text{ e}^{-\lambda \omega_\alpha} w[\phi|\phi_\alpha] \delta\left[\phi_\alpha  -\psi\right].
\end{eqnarray}
Eq.~(\ref{quantumtiltinggeneratorlevel1}), which has been derived by us with another method~\cite{Liu2021b}, is named the tilted Liouville-master equation in Hilbert space. Further defining an operator
\begin{eqnarray}
\label{tiltedrho}
\widetilde\rho(t)&=&\int D\psi D\psi^* \widetilde P[\psi,t] |\psi\rangle\langle \psi|,
\end{eqnarray}
in the previous work, we have shown that $\widetilde\rho(t)$ satisfies TQME~\cite{Liu2021b}:
\begin{eqnarray}
\label{tiltedquantumasterequation}
\partial_t \widetilde \rho(t)=-\text{i} [H,\widetilde\rho(t)]+\sum_{\alpha=1}^M r_\alpha\left ( \text{ e}^{-\lambda\omega_\alpha}A_\alpha\widetilde\rho(t)A^\dag_\alpha -\frac{1}{2}\left\{ A^\dag_\alpha A_\alpha,\widetilde\rho(t) \right \} \right ).
\end{eqnarray}
The last step of proving the equivalence is to substitute Eq.~(\ref{connectionofcountingsMP2PDP}) into~(\ref{tiltedrho}):
\begin{eqnarray}
\label{tiltedrho2}
\widetilde\rho(t)=\sum_{\alpha=1}^M\int_0^t q_\alpha(t,\tau)|U(\tau)[\phi_\alpha]\rangle\langle U(\tau)[\phi_\alpha]|d\tau.
\end{eqnarray}
We immediately find that the MGF of Eq.~(\ref{quantumcountingnumber}) achieves an alternative expression,
\begin{eqnarray}
\label{momentgenfuncquantumcounting}
M(\lambda,t)={\rm Tr}[\widetilde \rho(t)],
\end{eqnarray}
In fact, Eq.~(\ref{tiltedrho2}) also provides a more efficient way to derive TQME. The whole process is very similar to that of the MQME; see Appendix~\ref{AppendixA}.

{Now, we are in a position to explain the cause of separating Eq.~(\ref{quantumcountingnumber}) from the general Eq.~(\ref{countingnumber}): the proof of the equivalence definitely indicates that a closed equation such as the TQME does not exist for the general counting quantities. From this perspective, Eq.~(\ref{MEofcountingstatistics}) instead of TQME~(\ref{tiltedquantumasterequation}) has more potential in the general counting statistics. However, we admit that, thus far, such general statistics are not demanded in the communities of quantum optics and stochastic thermodynamics.}

Let us close the theoretical discussions by rewriting Eq.~(\ref{MEofcountingstatistics}) to a concise matrix equation:
\begin{eqnarray}
\label{MEofcountingstatisticsmatrixform}
{\bf G}\hat {\bf P}={\bf 1}_\delta,
\end{eqnarray}
where $M\times1$ matrix ${\bf \hat P}=(\hat P_1(v),\cdots,\hat P_M(v))^T$, ${\bf 1}_\delta=(\delta_{1\gamma},\cdots,\delta_{M\gamma})^T$, $\delta_{\alpha\gamma}$ is the kronecker symbol, the upper letter ($^T$) denotes transpose, and the diagonal and nondiagonal elements of the matrix ${\bf G}$ are
\begin{eqnarray}
({\bf G})_{\alpha\alpha}&=&\frac{1-\hat{p}_{\alpha|\alpha}(v)e^{-\lambda\omega_{\alpha\alpha}} }{\hat S_\alpha(v)}
\end{eqnarray}
and
\begin{eqnarray}
({\bf G})_{\alpha\beta}&=&-\frac{\hat{p}_{\beta|\alpha}(v)}{\hat{S}_\beta(v)}e^{-\lambda\omega_{\beta\alpha}},
\end{eqnarray}
$\alpha\neq\beta$, respectively. We have used the Laplace transform of Eq.~(\ref{survivalprob1}):
\begin{eqnarray}
\label{identitysMPs}
v\hat S_\alpha (v)+\sum_{\beta=1}^M \hat p_{\alpha|\beta}(v)=1.
\end{eqnarray}
Note that in the quantum case, the waiting time densities $p_{\alpha|\alpha}(\tau)$ are no longer forbidden. Using the inverse matrix of ${\bf G}$, we can solve the Laplace transform of the MGF:
\begin{eqnarray}
\label{LaplacetransformMGFmatrixform}
\hat M(\lambda,v)={\bf 1}{\bf G}^{-1}{\bf 1}_\delta,
\end{eqnarray}
where the $1\times M$ matrix ${\bf 1}=(1,\cdots,1)$. {We must emphasize that the size of the matrix $\bf G$ is $M\times M$, where $M$ is the number of collapsed wavefunctions and is usually equal to or less than dimension of a quantum system.}

\section{Resonant two-level quantum system}
\label{section6}
We use a simple open quantum system to illustrate the results obtained in the previous sections: a two-level atom is surrounded by an environment with finite inverse temperature $\beta$ and is driven by a resonant field. The MQME of the system in the interaction picture is
\begin{eqnarray}
\label{QMEtwolevel}
\partial_t\rho(t)&=&-{\rm i}\left[ H,\rho(t)\right ] + r_-[\sigma_-\rho(t)\sigma_+ -\frac{1}{2}\{\sigma_+\sigma_-, \rho(t) \}   ]\nonumber \\
&&+r_+[\sigma_+\rho(t)\sigma_- -\frac{1}{2}\{\sigma_-\sigma_+, \rho(t) \}   ].
\end{eqnarray}
Here, $H=-\Omega(\sigma_- +\sigma_+)/2$ represents the interaction Hamiltonian between the system and the resonant field, $\sigma_\pm$ are the raising and lowering Pauli operators, $\Omega$ is the Rabi frequency, and $r_\pm$ are the pumping and damping rates, respectively. Note that these two rates satisfy the detailed balance condition: $r_-=r_+\exp{(\beta\omega_0 )}$, and $\omega_0$ is the energy level difference of the two-level system. We set the two-level system to start with the ground state $|g\rangle$. The model is widely used in quantum optics~\cite{Mollow1975,Mandel1995,Scully1997} and quantum thermodynamics~\cite{Szczygielski2013,Alicki2018}.

There are two collapsed wavefunctions in the set: $\{\phi_1=|g\rangle,\phi_2=|e\rangle\}$. First, we verify Eq.~(\ref{steadystaterhoprobform}). To this end, we solve for the stationary rate $c_\alpha$, $\alpha=1,2$. The core of the calculations is the waiting time densities of the sMP embedded in the two-level quantum system, $p_{\alpha|\beta}(\tau)$, $\alpha,\beta=1,2$, in which Eq.~(\ref{waitingtimedensityquantum}) and the non-Hermitian Hamiltonian
\begin{eqnarray}
\label{nonHermitianHamiltonian}
\hat H=-\frac{\Omega}{2}(\sigma_-+\sigma_+) -\frac{\rm i}{2}r_+\sigma_-\sigma_+-\frac{\rm i}{2}r_-\sigma_+\sigma_-.
\end{eqnarray}
are used. This is a direct but tedious process. Hence, some relevant formulas remain in Appendix~\ref{AppendixB}. The final result is
\begin{eqnarray}
\label{rhosstwolevel}
\rho(t=\infty)&=&\frac{P_{2|1}\int_0^\infty  e^{-i\hat H\tau}|\phi_1 \rangle\langle\phi_1 | e^{i\hat H^\dag \tau}d\tau + P_{1|2} \int_0^\infty  e^{-i\hat H\tau}|\phi_2 \rangle\langle \phi_2 |e^{i\hat H^\dag \tau} d\tau}{P_{2|1}\tau_1+P_{1|2}\tau_2}\nonumber \\
&=& \frac{1}{2}I-\frac{r\delta}{2\Omega^2+r^2}\frac{1}{2}\sigma_z+{\rm i}\frac{\Omega\delta}{2\Omega^2+r^2}\sigma_+-{\rm i}\frac{\Omega\delta}{2\Omega^2+r^2}\sigma_-,
\end{eqnarray}
where we define $r=r_-+r_+$ and $\delta=r_--r_+$. Eq.~(\ref{rhosstwolevel}) is consistent with the steady state solution of Eq.~(\ref{QMEtwolevel}). Compared with the conventional method, which simply sets the left-hand side of the MQME to zero and solves a matrix equation in the $\sigma_z$ representation, the utility of Eq.~(\ref{steadystaterhoprobform}) appears to be slightly more complex.

Second, we apply the sMP to investigate the general counting statistics of the two-level quantum system. We write Eq.~(\ref{MEofcountingstatisticsmatrixform}) in an explicit form:
\begin{eqnarray}
\label{matrixeqforcountingstat}
\left[
  \begin{array}{cc}
    \frac{1-\hat p_{1|1}(v)e^{-\lambda\omega_{11}}}{\hat S_1(v)} & -\frac{\hat p_{2|1}(v)e^{-\lambda\omega_{21}}}{\hat S_2(v)} \\
    -\frac{\hat p_{1|2}(v)e^{-\lambda\omega_{12}}}{\hat S_1(v)} & \frac{1-\hat p_{2|2}(v)e^{-\lambda\omega_{22}}}{\hat S_2(v)} \\
  \end{array}
\right]\left[
               \begin{array}{c}
                 \hat P_1(v) \\
                 \hat P_2(v) \\
               \end{array}
             \right]=\left[
               \begin{array}{c}
                 1 \\
                 0 \\
               \end{array}
             \right].
\end{eqnarray}
Solving this $2\times 2$ matrix equation is trivial, and we obtain the Laplace transform of the MGF:
\begin{eqnarray}
\label{momentgeneratingfun2levelfinitetemperature}
\hat M(\lambda,v)=\frac{{\hat S_1(v)}\left[1-\hat p_{2|2}(v)e^{-\lambda\omega_{22}}\right]+{\hat S_2(v)}{\hat p_{1|2}(v)e^{-\lambda\omega_{12}}} }{\left[ 1-\hat p_{1|1}(v)e^{-\lambda\omega_{11}}\right ]\left[1-\hat p_{2|2}(v)e^{-\lambda\omega_{22}}\right] -\left[\hat p_{2|1}(v)e^{-\lambda\omega_{21}}\right] \left[\hat p_{1|2}(v)e^{-\lambda\omega_{12}}\right] }.
\end{eqnarray}
We observe that all terms involved in Eq.~(\ref{momentgeneratingfun2levelfinitetemperature}) have clear probability means. {If quantum counting quantities are concerned with, e.g., heat with weights in Eq.~(\ref{weightsofheat}) below,   Eq.~(\ref{momentgeneratingfun2levelfinitetemperature}) will agree with that solved by Laplace transform of the TQME.  More detailed explanations of a special case are provided in Appendix~\ref{AppendixC}.}

\begin{figure}
\includegraphics[width=1\columnwidth]{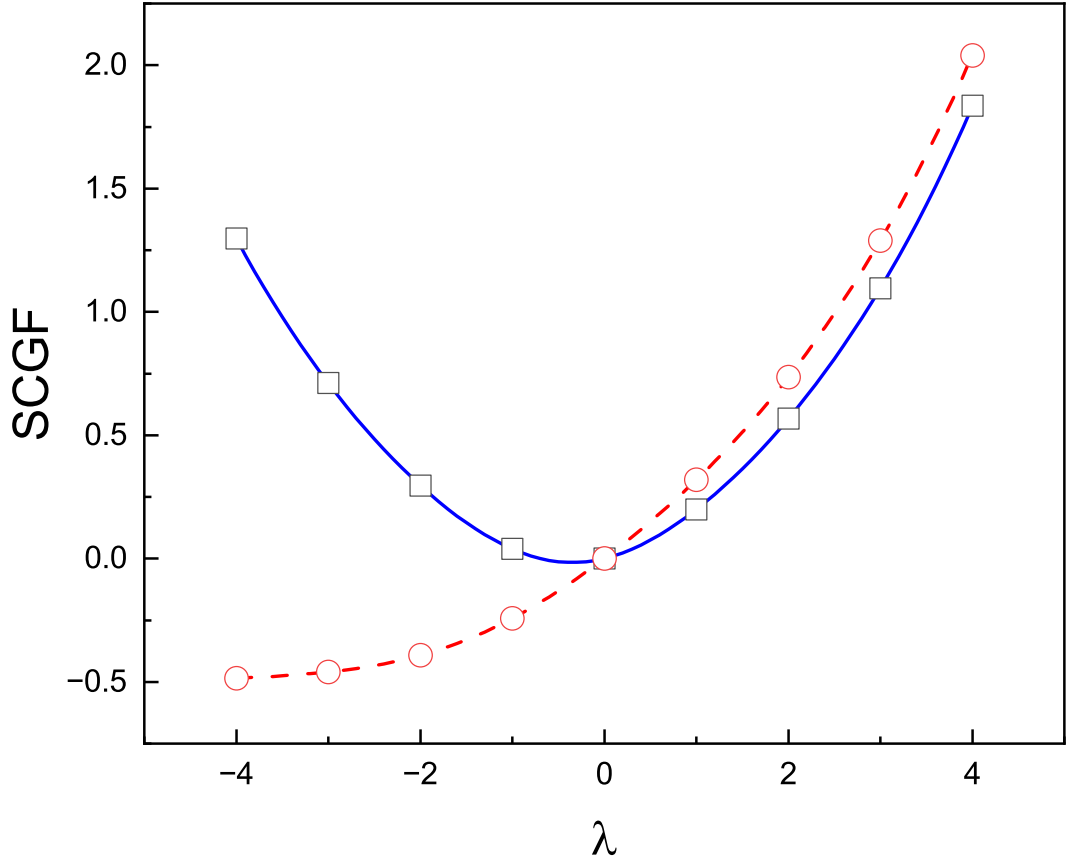}
\caption{The SCGFs data are solved by the sMP (curves) and TQME (symbols). The solid curve and squares are for the environment with finite temperatures, and the parameters are $r_-=1$, $r_+=0.5$, $\Omega=0.8$, and $\omega_0=1$. The dashed curve and circles are for the vacuum field, and the corresponding parameters are $r_-=1$, $r_+=0.0$, $\Omega=0.8$, and $\omega_0=1$.  }
\label{fig2}
\end{figure}

Eq.~(\ref{momentgeneratingfun2levelfinitetemperature}) is insightful in studying the fluctuation theorems~\cite{Lebowitz1999,Esposito2009,Liu2016a,Touchette2008}. Because of the detailed balance condition, we easily find that
\begin{eqnarray}
p_{2|2}(\tau)=p_{1|1}(\tau)e^{\beta\omega_0}.
\end{eqnarray}
Selecting physically relevant weights:
\begin{eqnarray}
	\label{weightsofheat}
\{\omega_{11},\omega_{12},\omega_{21},\omega_{22}\}\rightarrow\{-\omega_0,\omega_0,-\omega_0,\omega_0\}.
\end{eqnarray}
Then, Eq.~(\ref{quantumcountingnumber}) is the net energy released to the environment along quantum jump trajectories. From a thermodynamic viewpoint, this is the heat~\cite{Breuer2003,Horowitz2012,Liu2016a}. We can verify that the denominator of Eq.~(\ref{momentgeneratingfun2levelfinitetemperature}) is invariant under a transform $\lambda\rightarrow \beta-\lambda$:
\begin{eqnarray}
&&\left[ 1-\hat p_{1|1}(v)e^{\lambda\omega_0}\right ]\left[1-\hat p_{2|2}(v)e^{-\lambda\omega_0}\right] -\hat p_{2|1}(v) \hat p_{1|2}(v) \nonumber\\
&\rightarrow&{\left[ 1-\hat p_{1|1}(v)e^{(\beta-\lambda)\omega_0}\right ]\left[1-\hat p_{2|2}(v)e^{-(\beta-\lambda)\omega_0}\right] - \hat p_{2|1}(v)  \hat p_{1|2}(v)  }\nonumber \\
&=&{\left[ 1-\hat p_{2|2}(v)e^{-\lambda\omega_0}\right ]\left[1-\hat p_{1|1}(v)e^{\lambda\omega_0}\right] -\hat p_{2|1}(v) \hat p_{1|2}(v)  }.
\end{eqnarray}
Therefore, the poles of $\hat M(\lambda,v)$ are also invariant. According to the large deviation theory~\cite{Touchette2008}, the largest pole is the scaled cumulant generating function (SCGF) of the current $j=Q[X]/t$ in the long time limit~\cite{Andrieux2008},
\begin{eqnarray}
\varphi(\lambda)&=&\lim_{t\rightarrow\infty} \frac{1}{t}\ln M(\lambda).
\end{eqnarray}
Then, the following relationship holds:
\begin{eqnarray}
\label{GCsymmetry}
\varphi(\lambda)=\varphi(\beta-\lambda).
\end{eqnarray}
This indicates that the probability density of the current $j$ satisfies the fluctuation theorem in the steady state~\cite{Lebowitz1999}.

{The largest pole is obtained by vanishing the denominator of Eq.~(\ref{momentgeneratingfun2levelfinitetemperature}). In the case of heat, this implies  
\begin{eqnarray}
\label{largestpoletwolevelsystem}
\zeta^3+\zeta\left(4\mu^2-r_-r_+\right)-\frac{\Omega^2}{2}\left(r_-e^{-\lambda\omega_0}+r_+e^{\lambda\omega_0} \right)=0
\end{eqnarray}
with $\zeta=v+{r/2}$. 
This is a cubic equation and has an analytical solution: 
\begin{eqnarray}
\label{scaledgeneratingfun}
\varphi(\lambda)=-\frac{r}{2}+ \left[-\frac{x(\lambda)}{2}+\sqrt{ \Delta (\lambda) }\right]^{1/3}-\frac{y}{3}{\left[-\frac{x(\lambda)}{2}+\sqrt{ \Delta (\lambda)}\right]^{-1/3}},
\end{eqnarray}
where $\Delta(\lambda)={x^2(\lambda)}/{4}+{y^3}/{27}$, $x(\lambda)=\Omega^2({r_-e^{-\lambda\omega_0}+r_+e^{\lambda\omega_0}})/{2}$, and $y=4\mu^2-r_-r_+$. Eq.~(\ref{scaledgeneratingfun}) apparently satisfies the symmetry~(\ref{GCsymmetry}). The data are shown in Fig.~(\ref{fig2}) and are compared with those calculated by solving the largest eigenvalues of the TQME. We see that these two methods indeed present consistent data.

We mentioned that the TQME is inadequate if general counting quantities are considered. To illustrate this point, we define two random quantities, $Q_p[X]$ and $Q_n[X]$ with weights $\{1,0,0,1\}$ and $\{0,1,1,0\}$, respectively. Their meanings are apparent: the former is a counting of two consecutive collapses with the same wavefunctions, while the latter is a counting of two consecutive collapses with the distinct wavefunctions. Loosely speaking, current $j_p=Q_p/t$ indicates a frequency of resetting, while current $j_n=Q_n/t$ will indicate an activity of ``jumps" if we naively think of the quantum jump trajectories as a type of classical two-state jump process. Similarly, their SCGFs can be solved by looking for the largest poles of Eq.~(\ref{momentgeneratingfun2levelfinitetemperature}). Different from previous case of heat, 6-order algebraic equations are present. For instance, in $Q_n$ case we have  
\begin{eqnarray}
\zeta^2\left(\zeta^2+4\mu^2\right)^2 -\zeta \left(\zeta^2+4\mu^2\right)\left(r_-r_+e^{2\lambda}\zeta +\frac{r\Omega^2}{2}\right)	+\frac{r_-r_+}{4}\Omega^2\left(1-e^{2\lambda}\right)=0. 
\end{eqnarray}
Numerical schemes are needed and the data are shown in Fig.~(\ref{fig3}). Now, because the TQME is unavailable, in order to verify their correctness we have to simulate quantum jump trajectories. In inset of Fig.~(\ref{fig3}), we compare rate functions of large deviations~\cite{Touchette2008}, which are obtained by the simulation and doing Legendre transforms of the SCGFs, respectively. Their consistence is satisfactory. We also see that the ``jumps" are more active than the resetting, and their fluctuations are more significant. 
\begin{figure}
	\includegraphics[width=1\columnwidth]{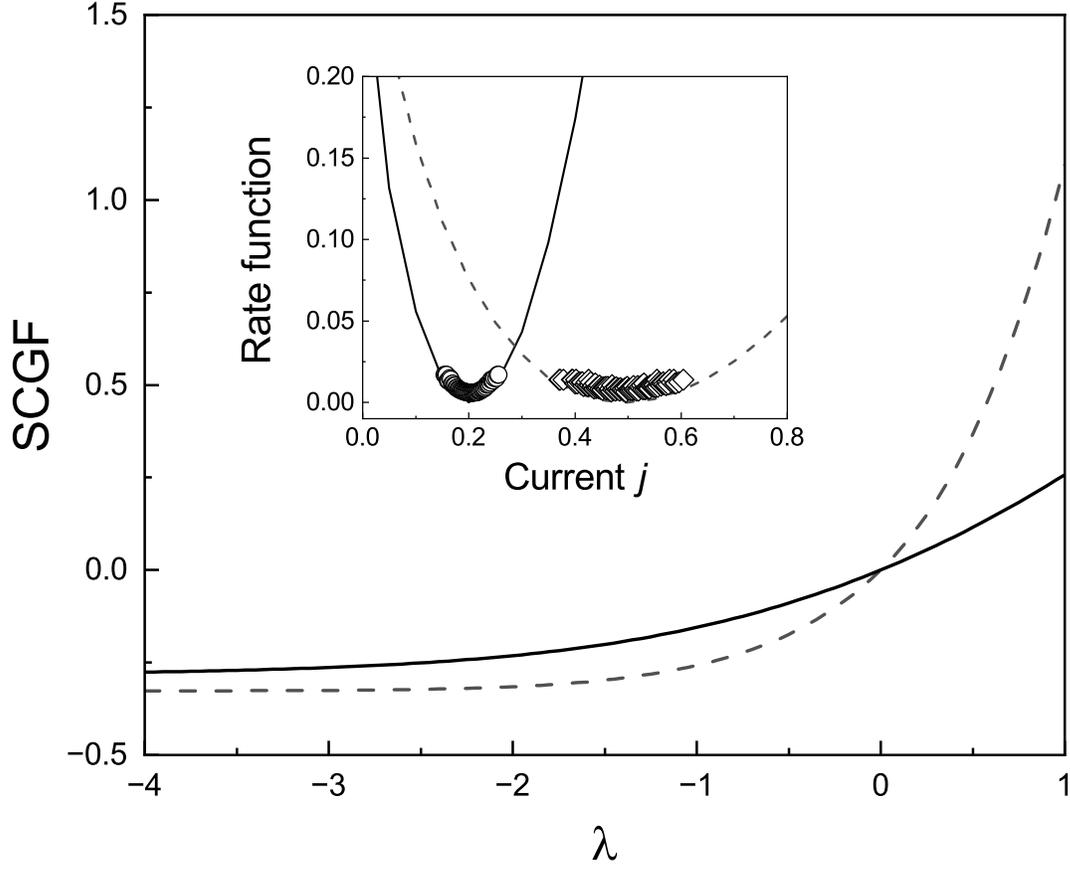}
	\caption{SCGFs of the random quantities $Q_p[X]$ (solid curve) and $Q_n[X]$ (dashed curve) solved by the sMP. Parameters are the same with those in the finite temperature case of Fig.~(\ref{fig2}). Inset shows the corresponding rate function of the current $j_l$ ($l=p,n$) obtained by simulations (symbols) and  performing Legendre transformations of the SCGFs (curves), respectively. The number of trajectories is 1000 trajectories. Simulation time is 5000.  }
	\label{fig3}
\end{figure}

}

\section{Conclusion}
\label{section7}
In this paper, we investigate the sMPs embedded in open quantum systems which are described by the Markov quantum master equations. With the assistance of the age-structure formalism, we prove that these stochastic processes inherit all statistical properties of the PDPs. Hence, the dynamics of the open quantum systems can be exactly reconstructed. Moreover, these embedded sMPs provide an alternative method to analyze and compute the counting statistics. This method is not only equivalent to TQME, which is now dominant in the literature, but also has more potential than TQME when the general counting statistics are considered. This is not surprising, since the foundation of the sMP is the PDPs, which possess more statistical information than the ``averaged" MQME. It will be interesting to investigate the sMPs embedded in more complex open quantum systems, e.g., electronic transport in nanosystems, in the near future.

\begin{acknowledgments}
This work was supported by the National Natural Science Foundation of China under Grant Nos. 12075016 and 11575016.

\end{acknowledgments}

\appendix
\section{Deriving MQME by Eq.~(\ref{densitymatrixzz2}) }
\label{AppendixA}
Taking the partial time derivative of Eq.~(\ref{densitymatrixzz2}) and substituting Eq.~(\ref{evoeqdoubletimeprob}), we have
\begin{eqnarray}
\label{derivativeofrho}
\partial_t\rho(t)&=&\sum_{\alpha=1}^M  p_\alpha(t,0)|\phi_\alpha\rangle\langle \phi_\alpha|-\sum_{\alpha=1}^M\int_0^t \Gamma_\alpha(\tau)p_\alpha(t,\tau) |U(\tau)[\phi_\alpha]\rangle\langle U(\tau)[\phi_\alpha]|\nonumber \\
&+&\sum_{\alpha=1}^M\int_0^t p_\alpha(t,\tau) \partial_\tau [|U(\tau)[\phi_\alpha]\rangle\langle U(\tau)[\phi_\alpha|]].
\end{eqnarray}
The first term of the right-hand side of Eq.~(\ref{derivativeofrho}) is a consequence of the integration of parts. Substituting Eq.~(\ref{doubletimeprobage0}) into it, we have
\begin{eqnarray}
\label{firstterm}
&&\sum_{\alpha=1}^M\sum_{\beta=1}^M\int_0^t p_\beta(t,\tau)k_{\beta|\alpha}(\tau)|\phi_\alpha \rangle\langle\phi_\alpha |d\tau \nonumber  \\
&=&\sum_{\alpha=1}^M\sum_{\beta=1}^M\int_0^t p_\beta(t,\tau)k_{\beta|\alpha}(\tau)
\frac{A_\alpha |U(\tau)[\phi_\beta] \rangle \langle [U(\tau)\phi_\beta] | A_\alpha^\dag   }{\parallel A_\alpha U(\tau)[\phi_\beta] \parallel^2}d\tau \nonumber \\
&=&\sum_{\alpha=1}^M r_\alpha {A_\alpha \rho(t)A_\alpha^\dag   },
\end{eqnarray}
where we have used Eqs.~(\ref{targetwavefunction}) and~(\ref{hazardfuncembeddedsMP}). Then,
using Eq.~(\ref{nonlinearSchrodingerequation}), we rewrite the third term to
\begin{eqnarray}
\label{thirdterm}
&&-{\rm i}\left[H,\rho(t)\right]-\frac{1}{2}\sum_{\alpha=1}^M r_\alpha \left \{ A_\alpha^\dag A_\alpha, \rho(t) \right \}+\int_0^t \sum_{\beta=1}^M r_\beta \parallel A_\beta|U(\tau)[\phi_\alpha]\rangle\parallel ^2 p_\alpha(t,\tau)|U(\tau)[\phi_\alpha]\rangle\langle U(\tau)[\phi_\alpha]|d\tau\nonumber\\
&=&-{\rm i}\left[H,\rho(t)\right]-\frac{1}{2}\sum_{\alpha=1}^M r_\alpha \left \{ A_\alpha^\dag A_\alpha, \rho(t) \right \}+\sum_{\alpha=1}^M\int_0^t \Gamma_\alpha(\tau) p_\alpha(t,\tau)|U(\tau)[\phi_\alpha]\rangle\langle U(\tau)[\phi_\alpha]|d\tau.
\end{eqnarray}
Substituting these two results into Eq.~(\ref{derivativeofrho}), we obtain the MQME, Eq.~(\ref{QME}).

\section{Some useful formulas in deriving Eq.~(\ref{rhosstwolevel}) }
\label{AppendixB}
Consider the case of $2\Omega>r$. We apply Eq.~(\ref{waitingtimedensityquantum}) to calculate the waiting time densities, $p_{\alpha|\beta}(t)$, $\alpha,\beta=1,2$. However, in fact, their Laplace transforms are more useful. Hence, here, we only list the latter:
\begin{eqnarray}
\label{Laplactransformpabtwolevelsystem1}
\hat p_{1|1}(v)&=&r_+\frac{\Omega^2}{2(v+r/2)\left[(v+r/2)^2+4\mu^2\right]} ,\\
\label{Laplactransformpabtwolevelsystem2}
\hat
p_{1|2}(v)&=&r_-\frac{(v+r/2)^2-\delta(v+r/2)/2+\Omega^2/2}{(v+r/2)\left[(v+r/2)^2+4\mu^2\right]}  ,\\
\label{Laplactransformpabtwolevelsystem3}
\hat p_{2|1}(v)&=&r_+\frac{(v+r/2)^2+\delta(v+r/2)/2+\Omega^2/2}{(v+r/2)\left[(v+r/2)^2+4\mu^2\right]}  ,\\
\label{Laplactransformpabtwolevelsystem4}
\hat p_{2|2}(v))&=&r_- \frac{\Omega^2}{2(v+r/2)\left[(v+r/2)^2+4\mu^2\right]} ,
\end{eqnarray}
where the parameters satisfy $16\mu^2+\delta^2=4\Omega^2$. We find that the two latter equations can be obtained from the former two equations by exchanging $r_-$ and $r_+$.

For the two-level quantum system, the stationary distributions of the Markov chain are simple:
\begin{eqnarray}
\label{pidistributiontwolevel}
\pi_1=\frac{P_{2|1}}{P_{1|2}+P_{2|1}},\hspace{0.5cm}\pi_2=\frac{P_{1|2}}{P_{1|2}+P_{2|1}};
\end{eqnarray}
see Eq.~(\ref{eigenvector}). The transition rates therein are
$P_{1|2}={\hat p}_{1|2}(0)$, $P_{2|1}={\hat p}_{2|1}(0)$, respectively. On the other hand, we can also make use of Eqs.~(\ref{Laplactransformpabtwolevelsystem1})-(\ref{Laplactransformpabtwolevelsystem4}) to calculate the mean times as follows:
\begin{eqnarray}
\tau_\alpha=-\frac{d}{dv}(\hat p_{\alpha|1}+\hat p_{\alpha|2})(0),
\end{eqnarray}
$\alpha=1,2$.
Finally, we solve
\begin{eqnarray}
\label{evolvingwavefunction1}
\int_0^\infty e^{-i\hat H\tau}|\phi_1\rangle\langle \phi_1 |e^{i\hat H^\dag\tau}d\tau
=\frac{2}{r(r^2+16\mu^2)}\left(
   \begin{array}{cc}
     2\Omega^2-r\delta+r^2 & {\rm i}\Omega(\delta-r) \\
     -{\rm i}\Omega(\delta-r) & 2\Omega^2 \\
   \end{array}
 \right),
\end{eqnarray}
and
\begin{eqnarray}
\label{evolvingwavefunction2}
\int_0^\infty e^{-i\hat H\tau}|\phi_2\rangle\langle \phi_2 |e^{i\hat H^\dag\tau}d\tau
=\frac{2}{r(r^2+16\mu^2)}\left(
   \begin{array}{cc}
    2\Omega^2 & {\rm i}\Omega(\delta+r) \\
     -{\rm i}\Omega(\delta+r) &  2\Omega^2+r\delta+r^2 \\
   \end{array}
 \right).
\end{eqnarray}
Substituting Eqs.~(\ref{pidistributiontwolevel})-(\ref{evolvingwavefunction2}) into Eq.~(\ref{steadystaterhoprobform}) and carrying out a careful simplification, we arrive at Eq.~(\ref{rhosstwolevel}).

\section{Solving the MGF and SCGF by the TQME }
\label{AppendixC}
We briefly describe the process of solving the MGF and SCGF by the TQME. {The quantum counting is the heat and its weights are given in Eq.~(\ref{weightsofheat}).} On one hand, this is for the convenience of the reader. On the other hand, through this process, we will acquire a direct understanding of the advantages and disadvantages of the embedded sMP and TQME.

First, we decompose the abstract operator $\widetilde\rho(t)$ as
\begin{eqnarray}
	\label{characteristicopertortwolevelsystem}
\widetilde\rho(t)=\frac{I}{2}\Sigma(t)+\frac{\sigma_z}{2}\Delta(t)+\sigma_+\bar{\sigma}_-(t)+\sigma_-\bar{\sigma}_+(t).
\end{eqnarray}
Define a $4\times 1$ matrix ${\bf q}(t)=(\Sigma,\Delta,\bar{\sigma}_-,\bar{\sigma}_+)^T$. In the $\sigma_z$ representation, the TQME, Eq.~(\ref{tiltedquantumasterequation}), has a matrix form:
\begin{eqnarray}
\label{matrixequationofTQME}
\partial_t {\bf q}(t)=[{\bf M}+{\bf B}(\lambda)]{\bf q}(t),
\end{eqnarray}
where
\begin{eqnarray}
 {\bf M}&=&\left[
   \begin{array}{cccc}
      0&  0 & 0 & 0 \\
     -r_-  + r_+ &  - r_-- r_+ & {\rm i}\Omega & -{\rm i}\Omega \\
     0 & {\rm i}\frac{\Omega}{2} & -\frac{r_-+r_+}{2} & 0 \\
     0 & -{\rm i}\frac{\Omega}{2} & 0 & -\frac{r_-+r_+}{2}  \\
   \end{array}
 \right],
\end{eqnarray}
and
\begin{eqnarray}
 {\bf B}(\lambda)&=&\left[
   \begin{array}{cccc}
     \frac{r_-}{2}\left( e^{\lambda\omega_0}-1\right) +\frac{r_+}{2}\left(e^{-\lambda\omega_0}-1\right)&  \frac{r_-}{2}\left(e^{\lambda\omega_0}-1\right) -\frac{r_+}{2}\left(e^{-\lambda\omega_0}-1\right) & 0 & 0 \\
     -\frac{r_-}{2}\left(e^{\lambda\omega_0}-1\right) +\frac{r_+}{2}\left(e^{-\lambda\omega_0}-1\right)&  -\frac{r_-}{2}\left(e^{\lambda\omega_0}-1\right) -\frac{r_+}{2}\left(e^{-\lambda\omega_0}-1\right)& 0 & 0 \\
     0 & 0 & 0 & 0 \\
     0 & 0 & 0 & 0  \\
   \end{array}
   \right],
\end{eqnarray}
respectively. Obviously, the matrix ${\bf B}(0)$ is zero. In this situation, Eq.~(\ref{matrixequationofTQME}) is nothing but the matrix equation of the MQME of the two-level quantum system, and the steady-state solution is Eq.~(\ref{rhosstwolevel}).

{ According to our theory, TQME~(\ref{matrixequationofTQME}) leads to the same MGF and SCGF as those of the sMP, Eqs.~(\ref{momentgeneratingfun2levelfinitetemperature}) and~(\ref{scaledgeneratingfun}). At first glance, this is not obvious. In addition, although the size of the matrix equation ($4\times 4$) is not too large, solving it by a manual way is a very tedious task and some software is resorted to. Therefore, we only address the simplest case that we can tolerate by a manual way, in which the atom is in a vacuum: that is, the pumping rate $r_+$ is zero. Taking the Laplace transform of Eq.~(\ref{matrixequationofTQME}) and solving the equation, we have
\begin{eqnarray}
\label{momentgenfunccounting2levelzerotemper}
\hat M(\lambda,v)&=&{\rm Tr}[\widehat {\widetilde \rho}(v)]=\frac{2v^2+3r_- v +  r_-^2 +2\Omega^2}
{2[2v ^3+3r_- v^2 +(2\Omega^2+r_-^2)v -(e^{-\lambda\omega_0}-1)r_-\Omega^2]} 
\end{eqnarray}
If the sMP method is used, Eq.~(\ref{LaplacetransformMGFmatrixform}) provides us
\begin{eqnarray}
\label{MGFtwolevelvacuum}
\hat M(\lambda,v)&=&\frac{ \hat S_2(v) }{1-e^{-\lambda\omega_0}\hat p_{2|2}(v)}\nonumber\\
&=&\frac{1-\hat p_{2|2}(v)}{v}\frac{1 }{1-e^{-\lambda\omega_0}\hat p_{2|2}(v)}.
\end{eqnarray}
The second equation is due to Eq.~(\ref{identitysMPs}). The reader is reminded that in this case only one collapsed wavefunction is present, that is, the size of matrix $\bf G$ is $1\times 1$. The differences between Eqs.~(\ref{momentgenfunccounting2levelzerotemper}) and~(\ref{MGFtwolevelvacuum}) are superficial: substitution of Eq.~(\ref{Laplactransformpabtwolevelsystem4}) into Eq.~(\ref{MGFtwolevelvacuum}) can verify their equivalence.

TQME calculates the SCGF by solving the largest eigenvalue of the matrix ${\bf M}+{\bf B}(\lambda)$. That is, we do not solve Eq.~(\ref{characteristicopertortwolevelsystem}). For the vacuum case, we find that the eigenvalues problem is equivalent to solving an algebraic equation:
\begin{eqnarray}
\label{eigenvalueeqution}
2\xi ^3+3r_-\xi^2 +(2\Omega^2+r_-^2)\xi -(e^{-\lambda\omega_0}-1)r_-\Omega^2=0 
\end{eqnarray}
where we set $\xi$ to the eigenvalue we are looking for. This is nothing but the pole of the denominator of Eq.(\ref{momentgenfunccounting2levelzerotemper}). Eq.~(\ref{eigenvalueeqution}) becomes simpler if we change variable $\xi$ to $\zeta=\xi+r_-/2$. Then, we have
\begin{eqnarray}
\zeta^3+4\mu^2\zeta-\frac{r_-\Omega^2}{2}e^{-\lambda\omega_0}=0.
\end{eqnarray}
This is a cubic equation and its real root is given by Cardano's formula. If the sMP method is used, the SCGF is obtained by setting the denominator of Eq.~(\ref{MGFtwolevelvacuum}) to zero: that is,
\begin{eqnarray}
\label{identitiySCGFtwolevel}	
\hat p_{2|2}(v)=e^{\lambda\omega_0}.
\end{eqnarray}
Of course, Eqs.~(\ref{eigenvalueeqution}) and~(\ref{identitiySCGFtwolevel}) are the same, but its probability meaning is not revealed until we obtain the latter.  

The above discussions have highlighted the respective advantages of the embedded sMPs and TQME. From a computational perspective, TQME is far superior to the sMP method. The former does not require any waiting time densities at all. In general, solving the matrix equation of the TQME or diagonalizing the equation for SCGFs are very simple when numerical schemes are used. Even so, we need emphasize that, if dimension of a quantum system is $D$, the size of the involved matrix is $D^2\times D^2$. In contrast, the size of the matrix $\bf G$ of the sMP method is $M\times M$ and $M$ is less than or equal to $D$. From a statistical and/or formal perspective, the sMP is more attractive, since all terms involved have clear probability means. In contrast, the TQME is abstract, and its matrix equation depends on the concrete quantum representations; its probability relevance is usually ambiguous.  

}


%

\end{document}